\documentclass[12pt,preprint]{aastex}

\newcommand{\myemail}{harris@cfa.harvard.edu}

\slugcomment{}

\shorttitle{}
\shortauthors{}

\begin{document}

\title{CAN MULTIBAND OBSERVATIONS CONSTRAIN EXPLANATIONS FOR KNOTTY JETS?}

\author{D. E. Harris}

\affil{MS-3, CfA, 60 Garden St.\\ 
Cambridge, MA 02138, USA\\
harris@cfa.harvard.edu}

\begin{abstract}
One can imagine a number of mechanisms that could be the cause of
brighter/fainter segments of jets.  In a sense, jets might be easier
to understand if they were featureless.  However we observe a wide
variety of structures which we call ``knots''.  By considering the
ramifications of the various scenarios for the creation of knots, we
determine which ones or which classes are favored by the currently
available multiwavelength data.
\end{abstract}


\section{Introduction}	

With the advent of the Chandra X-ray Observatory, sub arcsec
resolution became available in the 0.5 to 10 keV band.  This
technological advance resulted in an increase in the number of sources
with detections of X-ray jets and hotspots from a few to close to one
hundred.  During the last ten years, it became apparent
that the conventional wisdom (that knots in jets were the results of
internal shocks) was only part of the story.  We now realize that for
X-ray synchrotron features, any emission region must also be an
acceleration region since the timescales for radiative losses of the
emitting electrons is much shorter than travel times from one part of
the jet to downstream regions.  Characteristic loss times for X-ray
synchrotron-emitting electrons is ten times shorter than for those
responsible for the optical emission.

\subsection{Angular resolution vs. physical resolution}

One of the main obstacles in comparing features in FRI jets with those
in quasar and FRII jets is the large change in physical resolution
for a fixed angular resolution.  In Table~\ref{tab:res} we give the
physical resolution corresponding to one arcsec for a sample of
sources.  Typical angular resolutions for adaptive optics, the VLA,
MERLIN, and HST will be 0.1 arcsec and VLBA can provide resolutions of
1 to 10 milli arcsec.

\begin{table} 
\begin{center}
\caption{Physical size corresponding to one arcsec}
\begin{tabular}{llrrl}
\tableline\tableline

Source  & redshift  &    Distance  &  size  & Description\\ 
&  & (Mpc) & (pc) & \\ 
\tableline

Cen A  &     &  3.5   &   17  & FRI jet   \\ 
M87   &    0.00427  &   16    &    77   & FRI jet  \\ 
PicA   &   0.035   &   152    &   700    &  FRII jet and hotspot \\ 
CygA    &  0.056   &   247   &   1070  & FRII hotspots \\   
3C273  &   0.158  &    749   &   2700   & CDQ jet  \\ 
3C109  &   0.3056  &  3338   &   4500   &  FRII hotspot \\ 
3C263  &   0.656   &  3934   &   7000   &  LDQ hotspot \\ 
3C280  &   1      &   6600   &   8000   &  FRII hotspots \\ 
3C9    &   2      &  15850   &   8500   &  LDQ jet \\ 
1508   &   4.3    &  39809   &   6900   &  CDQ knot \\ 
\tableline
\end{tabular} \label{tab:res}
\end{center}
\end{table}

\begin{figure} 
\epsscale{0.8}
\plotone{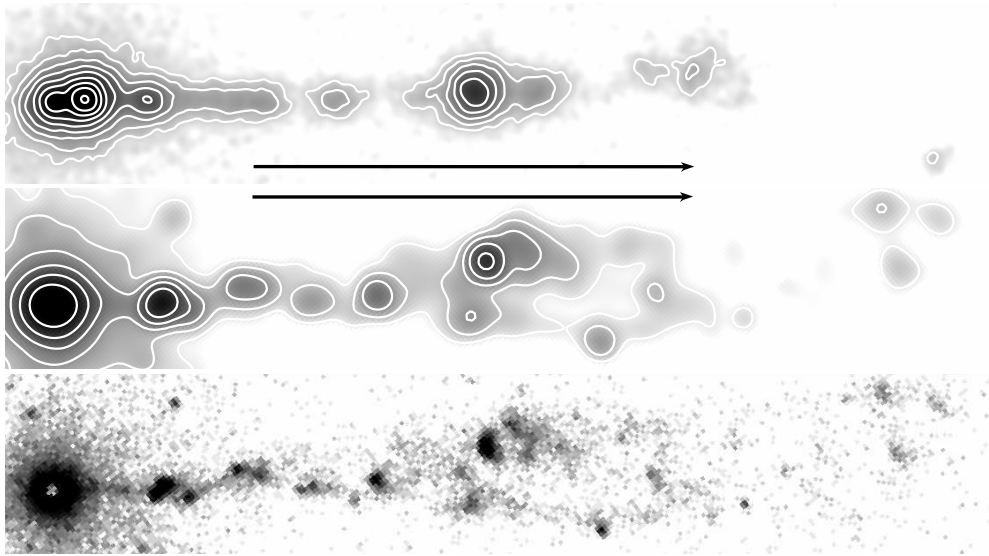}
\caption{Chandra images of the M87 jet (top) and the Cen A jet (middle
and bottom).  All images have been rotated and contours increase by
factors of two.  The scale bars are 1 kpc long.  The M87 image has
been regridded to one tenth of a native ACIS pixel and smoothed with a
Gaussian of FWHM=0.25$^{\prime\prime}$.  The center image is how the
Cen A jet would appear if it were at the distance of M87.  The bottom
image shows the event file which demonstrates that the lower resolution
image misses the fine scale structures. \label{fig:cena}}
\end{figure}

When we make our calculations for physical parameters of jet knots, we
are most likely making gross errors because of this limitation.  In
fig.~\ref{fig:cena} we can understand that measuring the size,
brightness, intensity, and spectrum of a knot in the M87 jet, or in 
the jet of Cen A (if it were at the distance of M87), will not provide the
data necessary to derive the correct physical parameters for the features we
can actually discern in the full resolution X-ray map (bottom panel).

\subsection{Is the X-ray emission from quasar jets synchrotron or inverse 
Compton emission ?}

While no definitive answer to this question has been demonstrated, we
are reasonably confident that X-ray emission from FRI jets is
dominated by synchrotron emission.  The remaining contentious issue is
the nature of the X-ray emission from quasar jets.  If it is
synchrotron, then the electrons responsible for the radiation will
have energies of $\gamma\geq~10^7$ ($\gamma$ is the Lorentz factor of
the electrons).  However, since the energy density of the cosmic
microwave background (CMB) in the jet frame will be augmented by
$\Gamma^2$ ($\Gamma$ is the bulk Lorentz factor of the jet) if
$\Gamma\geq$5, the resulting inverse Compton (IC) emission comes from
electrons with $\gamma\approx$100.  In the former case $E^2$ losses
(i.e. synchrotron and IC) result in halflives of order a year, whereas
in the latter case, it would be $\geq~10^5$ years, and even longer for
regions with magnetic field strengths significantly less than 1
mG~\citep{hk02}.  Knot morphology and intensity ratios between radio and
X-ray might be quite different depending on which process dominates
the X-ray emission.

Synchrotron self-Compton (SSC) emission (for which the target photons 
are the synchrotron photons) has been found to provide reasonable
fits to the spectra of some FRII hotspots, but has generally not been 
able to explain knot emission~\citep{hard04}.

\section{Multiband Aspects of Knots} 

For almost all X-ray detections of knots and hotspots, there is a very good
correspondence between the X-ray and radio morphologies, but the intensity
ratio (X-ray flux divided by the radio flux) varies considerably.

\subsection{The X-ray to radio intensity ratio, $R_{\rm{xr}}$}

From a study of over a hundred knots and hotspots with both radio and
X-ray detections, it has been found that $R_{\rm{xr}}$ lies in the range 1 to
100 for knots of both FRI radio galaxies and quasars whereas most hotspots (both
FRII and quasar) have ratios in the range 0.03 to 3~\citep{harr10}.  In
that work we did not sample the much larger number of radio knots with
only upper limits of X-ray intensity, so it is quite likely that there
are many knots and hotspots with smaller $R_{\rm{xr}}$ values.  {\it A
priori}, if quasar knots were all dominated by IC/CMB X-ray emission,
we might have expected to see a clear difference in $R_{\rm{xr}}$ between
quasars and FRI knots; instead we find essentially the same range for
both classes of sources.

\subsection{Offsets and progressions}

``Offsets'' is a term we use to describe a common (but not universal)
property of individual X-ray knots.  In many cases, when observed with
similar angular resolutions, the brightness of the X-ray image peaks
upstream of the lower frequency emissions.  This behavior is seen also
when comparing optical and radio morphologies, and to the best of our
knowledge, always occurs in the sense that the higher frequency peaks
upstream of lower frequencies.  This subject is discussed in section
3.2.1.1 of \citet{hk06}, and several examples are shown.

The term ``progressions'' is used to describe a systematic change in the
overall spectral properties of knots as a function of distance from
the nucleus.  This has also been covered in \citet{hk06}
(section 3.2.1.2).  The systematic change is best seen in the ratio of
X-ray flux to radio flux and occurs in the sense that the ratio is
larger closer to the nucleus.  In the case of 3C~273, the ratio
changes by two orders of magnitude, whereas for 4C~19.44, the effect
is essentially absent.

It is not difficult to see that if a jet segment that behaved like
3C~273 were observed with a single resolution element, it would
show a marked offset between the peak of the X-ray compared
to the peak of the radio distribution.

Since this sort of effect has been observed over many physical scales,
it is likely that synchrotron loss time compared to travel time down
the jet is not the only cause of offsets.  Rather some other mechanism
is at work such as a progressively larger field strength moving down
stream.  That would produce higher radio intensities as well as
perhaps curtailing the production of the very high energy electrons
required to produce X-ray synchrotron emission.


\section{Mechanisms for Knot Production} 

Our basic assumption is that a knot is a region of enhanced emissivity
which is produced by the jet.  It is moving relativistically, but not
necessarily at the same velocity as ``the jet'' (i.e. the velocity of
the power flow)~\citep{hk07}.  One can imagine several mechanisms for
modulating the emissivity along a jet.  We consider several
possibilities, and suggest a few diagnostics.  We do not consider the
underlying reasons for the existence of any particular knot at any
particular location (i.e. instabilities, interactions with stars,
molecular clouds, etc).

\subsection{The classical shock scenario}

Perhaps the most intuitive explanation for knotty jets is the common
notion of a series of shocks.  Each would create a new supply of
relativistic electrons with a power law distribution determined by the
local conditions.  The eventual dimming as the shocked plasma is
advected down stream can be caused by $E^2$ losses or expansion (first
power of energy).  For $E^2$ losses, we expect the lower frequency
emissions to last longer, leading to offsets in peak brightness as we
move downstream.  This is often seen in synchrotron jets; for IC/CMB
models, the X-rays come from low energy electrons and should last
longer.  This behavior is almost never seen, although the end of the
jet in 4C19.44 could be an example.

\subsection{Adiabatic expansion/contraction}

Although quite similar to the shock scenario, there is no shock
acceleration as such.  The only changes to the electron energy
distribution comes from the change in volume of the emitting region.
In both the shock case and the change in volume, compression augments
the magnetic field and boosts the energy of electrons; expansion reduces
synchrotron emission both from the lowering of electron energies, but
also by the drop in field strength and moreover, for a fixed observing
band, a lower field means you are observing higher energy electrons
than previously so you are sampling a segment of the electron spectrum
that has many fewer electrons.  In the case of IC/CMB X-rays, the
emissivity drops only because the normalization factor of the power
law distribution of electron energy drops; the change in magnetic field has
no effect.  Therefore, if expansion were to be the dominent operator
for separating adjacent knots, it would mean that the contrast from
knot to inter-knot should be greater in the radio/optical than in the
X-rays for the IC/CMB model whereas if synchrotron emission dominates
the X-ray emissivity, the contrast should be sensibly the same at
all frequencies.

\subsection{Episodic activity/ejection - power flow is not constant}

If kpc jets are similar to pc scale jets, an episodic supply of power
to the jet by the SMBH could produce a series of moving knots: knots
represent high power intervals of activity, gaps are when the power is
low or absent (c.f. the ``flip-flop'' model of jet formation).  If
such a mechanism were to be the only formative one, there would have
to be two timescales: one for pc scale jets and the other for kpc
scale jet knots.  Current evidence does not favor this scenario,
e.g. the upstream edge of HST-1 (a jet knot close to the nucleus of
M87) was thought to have an apparent velocity close to c (downstream
blobs were estimated at 6c~\citep{bir99}) from HST data in the 1990's.
More recently, we measured comparable values at 1.7 GHz~\citep{che07}.
However, the upstream edge of HST-1 has not moved during the
intervening 10 years, consistent with an interpretation in terms of a
stationery shock. We suspect that both estimates for the motion of the
upstream edge were centroid shifts caused by the ejection of new
components.

\subsection{Doppler boosting along a curved trajectory}

If the path of a jet changes direction compared to the line of sight,
either by thrashing or by a regular (e.g. helical) path, apparent
knots can be produced even though the jet itself has a steady power
flow. Once the jet has $\Gamma~\geq$ a few, moving in and out of the
beaming cone can produce the required brightness fluctuations.

The simple expectation is that the radio and X-ray emissions will be
coincident: each knot will have the same location and morphology for
all wavelengths, i.e. no offsets are expected in the brightness
distributions.  If the X-rays come from IC/CMB, for most jets the
contrast should be higher for the X-rays because of the extra beaming
factor of IC/CMB ~\citep{hk02,mas09}.

\subsection{Variable Beaming Factor}

If it were possible to ``store'' jet energy in some other form than in
the bulk Lorentz factor, it might be conceivable to envisage a jet
with an oscillating value of $\Gamma$.  However, if the total energy
flow is proportional to $\Gamma$, it would seem difficult to allow
$\Gamma$ to drop substantially and then to increase again.  If we 
separate the emitting region from the underlying jet, then the knot's
emission might well decay from a drop in $\Gamma$, and the subsequent
knot would have to rely on one of the other possible explanations to
generate a new emitting volume.

\section{Summary} 

What we observe is not necessarily ``the jet''.  ``The jet'' is whatever
it is that carries the energy from the environs of the SMBH to
distances of hundreds of kpc.  There are a number of possibilities:
magnetic field/ Poynting flux, hot or cold protons, or cold pairs.
What we see are hot electrons/positrons, but these cannot be the agent
that transports the energy since there are inescapable IC losses for
electrons with $\gamma\geq$2000~\citep{hk07}.  The larger the $\Gamma$ of
the jet (to minimize the local flow of time for $E^2$ losses), the
larger the IC losses since the effective energy density of the CMB
increases as $\Gamma^2$.  Thus we have the river analogy: ``the jet''
is a river with smoothly flowing water; the emission we see is like
white water produced by turbulence around rocks in the river or
waterfalls.  The white water is a product of the river and may well be
carried along by the river's flow, but not necessarily with the
underlying velocity of the water.  All our observations describe the
product, not the jet itself.  When we see a knot, we see a location
where energy is transferred from the jet to produce hot (radiating)
electrons.  For many knots, the energy transferred is a small fraction
of the total power of the jet, whereas for terminal hotspots in FRII
radio galaxies, the transfer is complete.

If IC/CMB dominates the X-ray emission, it seems that creating knot
structure is a non-trivial problem.  Curved trajectories and episodic
ejection from the SMBH may be amongst the few viable options, since
once a substantial population of low energy electrons is generated, it
is difficult to reduce the emissivity and then increase it again.

We suggest a diagnostic of comparing the brightness contrast between
the peak intensity of a knot and the preceding and following minimum
brightness (the gap between knots).  If expansion and contraction is
the dominant mechanism for knot production, the synchrotron emission
should have a much larger contrast than IC/CMB emission.  If however a
change in direction of the beaming cone is the principal operative,
then the IC/CMB emission should (statistically) display the higher
contrast.

\section*{Acknowledgments} 

It is a pleasure to acknowledge collaborators C. C. Cheung,
F. Massaro, and L. Stawarz.  Partial support for this work was
provided by NASA grants AR6-7013X and G09-0108X.


\end{document}